# Research on diffusion of Mo substrate atoms into Ti and Cr thin films by secondary ion-ion emission method

A. D. Abramenkov, Ya. M. Fogel', V. V. Slyozov, L. V. Tanatarov, O. P. Ledenyov

*National Scientific Centre Kharkov Institute of Physics and Technology, Academicheskaya 1, Kharkov 61108, Ukraine.*

The experimental research on the nature of diffusion by the *Mo* substrate atoms into the *Ti* and *Cr* deposited thin films is completed by the secondary ion-ion emission method. In [1], the initial stage of the *Ti* thin film on the *Mo* substrate deposition process, using the *Ti* evaporation technique in the vacuum, is researched. It was found that the *Mo* substrate atoms diffuse into the continuously deposited *Ti* thin film. The diffusion of *Mo* substrate atoms by the nodes of crystal grating in the deposited metallic *Ti* thin film with the continuously increasing thickness is theoretically considered in [2]. In this research, the diffusion coefficients of *Mo* substrate atoms into the *Ti* and *Cr* thin films are measured by the secondary ion-ion emission method in the *Mo-Ti* and *Mo-Cr* systems.

PACS numbers: 79.60.Jv, 79.90.+b, 68.35.Fx, 68.35.bd, 68.55.-a, 68.60.-p, 66.30.-h, 66.70.Df, 61.80.Jh.

Keywords: Molybdenum (*Mo*) substrate, Titanium (*Ti*) and Chrome (*Cr*) deposited thin films, diffusion of atoms, diffusion coefficients, secondary ion-ion emission method.

## Introduction

In [2], it was shows that in the two layered system substrate − deposited thin film (*DTF*), the concentration $C$ of substrate's atoms at the boundary between the *DTF* and the vacuum at the big enough thicknesses of the *DTF* is expressed as in the time dependent eq. (*1*)

$$C(t) = At^{-\frac{3}{2}} \exp[-\frac{k^2}{4D}t], \quad (1)$$

where $D$ is the diffusion coefficient of substrate's atoms into the material of the *DTF*, $t$ is the time; $A$ is the constant.

Applying the secondary ion-ion emission method [1, 3], it is possible to register the changing of concentration magnitude of the *Mo* substrate's atoms, diffused through the *Ti* or *Cr* *DTFs* with the continuously increasing thickness over some time period $t$. Since the second ions current $I$, corresponding to the *Mo* substrate's atoms, which are situated on the *Ti*- or *Cr-vacuum* surface, is proportional to the concentration of *Mo* atoms, hence the dependence of the second ions current of the $Mo^+$ on the time $I(t)$ can be described by the formula (*1*). Going from this consideration, a new measurement method of diffusion coefficient's magnitude $D$ in the solid states can be proposed. The essence of this method consists in the following things:
1) the research of dependence $I(t)$ for the ions of substrate's substance;
2) the finding of value $k^2/4D$, using the tangent of the angle of tilt in the dependence $\ln I \cdot t^{3/2} == f(t)$; and
3) the measurement of velocity of growth of thin film thickness.

## Research on diffusion of *Mo* substrate atoms into *Ti* and *Cr* thin films by secondary ion-ion emission method

In the present research, the diffusion coefficients of the *Mo* substrate's atoms in the *Ti* and *Cr* thin films are measured by the secondary ion-ion emission method during the deposition on the *Ti* and *Cr* thin films on the *Mo* substrate. Since the *Ti* and *Cr* metals together with the *Mo* metal create the solutions of substitution, hence it is possible to use the formula (*1*) to describe the diffusion process in these metals. In [1], the measurement set up, and the method of mass-spectrometry research on the diffusion of the substrate's atoms in the deposited thin film with the use of phenomena of secondary ion-ion emission, are described. The primary argon ions $Ar^+$ were used to irradiate the metal-vacuum surface in this research.

In Fig. 1, the dependence of the current of the secondary ions $Mo^+$ on the time $I(t)$ for the *Mo - Ti* system at the temperature of *1000 °C* is presented. This dependence was obtained during the continuous deposition of the *Titanium* thin film with the velocity of *0.5* monolayer per minute on the *Molybdenum* ribbon at the temperature *T=1000°C*. The computed curve $I(t)$, calculated with the use of eq. (*1*), is also shown in Fig. 1. As it can be seen, the experimental curve of the dependence $I(t)$ is well approximated by the computed curve at the big enough thicknesses of the *Titanium* thin film. The similar correlation is observed in the case of *Molybdenum- Chrome* system.

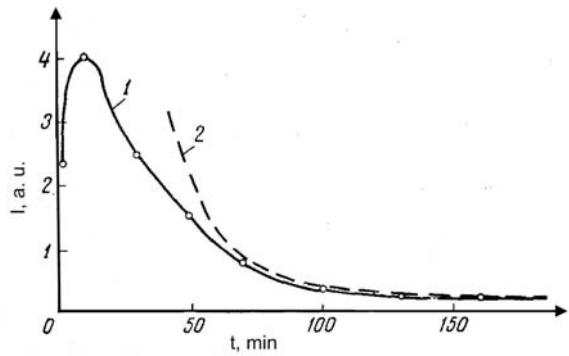

*Fig. 1. Dependence of current of secondary ions Mo$^+$ on time I(t) for Mo - Ti at temperature of 1000 °C:*
*1 - experimental curve; 2 – computed curve, calculated with the use of eq. (1).*

The value of expression $k^2/4D$ was calculated, using the obtained tangent of the angle of tilt in the dependence $\ln I \cdot t^{3/2} = f(t)$, created for the time period, when the experimental and computed curves $I(t)$ were coincided. The deposition velocity of the $Ti$ and $Cr$ thin films was determined by the weighting of the layer of thin film's material, deposited per certain time. The diffusion coefficients of the $Mo$ substrate's atoms in the deposited $Ti$ and $Cr$ thin films were determined with the help of the above described two measurements.

The diffusion coefficients of the $Mo$ substrate's atoms in the deposited $Ti$ and $Cr$ thin films were measured at a number of the temperatures $T$ of the $Mo$ substrate. The same $Mo$ substrate was used to perform a series on measurements. Every subsequent experiment was performed, when the deposited $Ti$ or $Cr$ thin film layers, created on the surface of the $Mo$ ribbon at the previous experiment, was evaporated due to the $Mo$ ribbon heating up to the temperature of $1400°C$. The value of the diffusion coefficient $D$ was determined as an average value among a number of the values, obtained in several experiments. The deviation among the values of the diffusion coefficient $D$ was in the range of 10-12%.

In Fig. 2, the dependences $\ln I \cdot t^{3/2} = f(t)$ are created, using the results of diffusion coefficients measurements during the diffusion by the $Mo$ atoms into the $Ti$ and $Cr$ deposited thin films. As it can be seen in Fig. 2, the dependences $\ln I \cdot t^{3/2} = f(t)$ are linear in the cases of the $Mo$-$Ti$ and $Mo$-$Cr$ systems. Going from the tangent of the angle of tilt of the direct lines in Fig. 2, the activation energies of diffusion by the $Mo$ substrate atoms into the $Ti$ and $Cr$ thin films are calculated. The values of activation energies are *2.76 eV* ($Mo$-$Ti$) and *2.9 eV* ($Mo$-$Cr$). The diffusion coefficients for the $Mo$-$Ti$ and $Mo$-$Cr$ systems can be calculated, using the following formulas

$$D = 5,3 \cdot 10^{-3} e^{-\frac{64000}{kT}} \; ; \; D = 1,6 \cdot 10^{-2} e^{-\frac{67000}{kT}} \, .$$

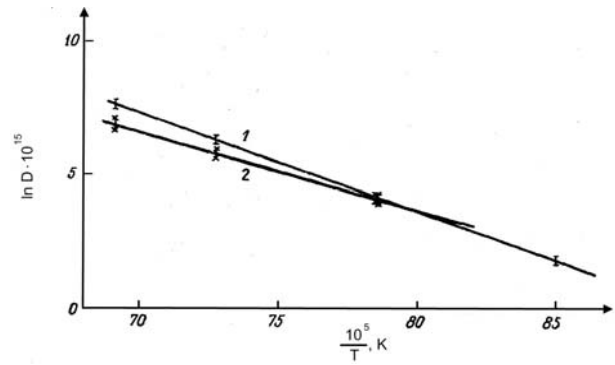

*Fig. 2. Dependence of lnD on 1/T for Mo - Ti (curve 1) and Mo - Cr (curve 2).*

In the case of the diffusion of the $Mo$ substrate atoms into the $Ti$ thin film, the measurement of diffusion coefficient was completed, using the $Mo^{99}$ radioactive isotope in [4]. At the diffusion of the $Mo$ substrate atoms into the $Ti$ thin film, the value of diffusion coefficient is *1.7·10$^{-9}$ cm$^2$/sec*, and the activation energy is *1.4 eV*. The big difference between the present research results and the early reported research [4] can be attributed to the fact that the diffusion of the $Mo$ substrate atoms into the $Ti$ thin film was mainly realized through the defects (the grain boundaries, dislocations) in the $Ti$ thin film's crystal structure in [4]. At the deposition of the $Ti$ thin film on the $Mo$ substrate by the evaporation technique in the vacuum, the created $Ti$ thin film has a small concentration of defects [5], and the diffusion by the $Mo$ substrate atoms into the $Ti$ thin film has place through the crystal grating. As far as the $Mo$-$Cr$ system is concerned, it is necessary to emphasis that the experimental results, obtained by the secondary ion-ion emission method, are in good agreement with the data in [6, 7].

## Conclusion

The experimental research on the nature of diffusion by the $Mo$ substrate atoms into the $Ti$ and $Cr$ deposited thin films is completed with the application of the secondary ion-ion emission method. The diffusion coefficients of $Mo$ substrate atoms into the $Ti$ and $Cr$ thin films are measured. The comparative analysis on the obtained research results is provided.

This research paper was published in *The Physics of Metals and Metallography* in 1970 [8].

*E-mail: ledenyov@kipt.kharkov.ua

---